\documentclass[aps,prb,twocolumn,showpacs]{revtex4}
\usepackage{graphicx}
\begin{document}

\title{Magneto-transport and divergent screening of driven two dimensional electron gas \\ in high Landau levels}

\author{A. Kashuba}

\affiliation{L. D. Landau Institute for Theoretical Physics, Russian Academy of Sciences, 2 Kosygina str., 119334 Moscow}

\date{\today}

\begin{abstract}
In two dimensional electron system in magnetic fields such that the Fermi energy lies in high Landau levels and driven by microwave $ac$-field a divergence of the Coulomb screening due to the Ryzhii photocurrent is predicted. A model of magneto-transport with superposition of long range and short range disorders is introduced. In this model the larger is the gradient of the long range potential the smaller is the longitudinal conductivity/resistivity. As a consequence a qualitative theory of the recently discovered zero-resistance states is given. 
\end{abstract}


\maketitle

\section{Introduction}

Electron systems interacting with microwaves could be driven into highly non-equilibrium state with unusual properties. For example two dimensional electron systems (2DEG) in GaAs heterostructures subjected to a weak perpendicular magnetic field show microwave induced resistance oscillations which in the limit of extreme power are capped by zero resistance states \cite{Mani,zdpw}. Ryzhii has predicted that an additional electric current will flow against or along applied electric field in the 2DEG driven by microwaves depending on the ratio of microwave frequency $\omega$ and the cyclotron frequency $\omega_H$. \cite{ryzh} Recent works \cite{dsrg,ll03} have clarified the origin of the Ryzhii current and resistance oscillations in disordered systems. At strong microwave driving the negative Ryzhii current could overcome usual positive dissipative current and thus results in the total local negative conductivity. Negative conductivity leads to the current domain instability and the domains could arrange themselves to give a zero total resistance \cite{aam03}. 

Zero resistance states may have different origins. For slow driving $\omega\ll \omega_H$ a wide suppression of resistance \cite{dsuk04} could be a manifestation of 'driven' degeneracy of electron states \cite{kash04}. Fast driving $\omega\gg \omega_H$ averages out the disorder potential and would lead to an ideal state with zero resistance (not reported yet to my knowledge), similar to the induced transparency in optics. This paper is concerned with the intermediate frequencies $\omega\sim \omega_H$, where multiple zero resistance states specific only to 2DEG in magnetic field emerge due to the Ryzhii phenomenon \cite{Mani,zdpw}.

In this paper I study effects of the long range disorder due to the Si-doped layer in high mobility heterostructures. If a long range potential is present in the electron system driven by microwaves then random Ryzhii currents will be locally induced even in the absence of the probe $dc$-voltage. Electrons will move away from the equilibrium density distribution, thus, modifying the Coulomb screening. The relationship between the bare long range potential $u_0(\vec{r})$ and the screened potential $u(\vec{r})$ is linear one: $u(\vec{q})= F(\vec{q}) u_0(\vec{q})$, where $F(\vec{q})$ is the screening factor, provided both $|u|,|u_0| \ll \epsilon_F$. One of the central results of this paper is that:
\begin{equation}\label{screening}
F^{-1}(\vec{q})= 1+\frac{2\pi e^2}{|\vec{q}|} \nu(\epsilon_F) \left(1- \frac{\sigma_R(W)}{\sigma_D(W)} \right)
\end{equation}
where $\sigma_D(W)$ is a diffusion part of the total conductivity $\sigma(W)=\sigma_D(W)- \sigma_R(W)$ and $\sigma_R(W)$ is the Ryzhii conductivity (assuming that $k\omega_H< \omega<(k+1/2) \omega_H$, where $k$ is an integer). They both depend on the intensity of microwaves $W$. In the long range limit $qR_c\ll 1$, where $R_c$ is the cyclotron radius, the electron transitions amount to a continuous current which is the linear response to two different perturbations: weak external electric field and weak gradient of electron density:
\begin{equation}\label{totcur}
e\vec{j}(\vec{r})=\sigma(W)\vec{\nabla} u(\vec{r})+e^2D(W)\vec{\nabla} n(\vec{r}),
\end{equation}
where $D(W)$ is the diffusion constant. The Einstein relationship states the detail balance at equilibrium: $\sigma(0)=e^2\nu(\epsilon_F)D(0)$ ($\sigma_R(0)=0$). It is violated in the driven state with $\sigma_D(W)=e^2\nu(\epsilon_F)D(W)$. The source of electrostatic potential $u(\vec{r})$ (in this paper it means the electrostatic energy) is either donors or electrons:
\begin{equation}\label{screen}
u(\vec{q})=u_0(\vec{q})+(2\pi e^2/|\vec{q}|)\ n(\vec{q}).
\end{equation}
The total current under the stationary microwave power is zero $\vec{j}(\vec{q})=0$. Assuming that electron relaxation due to phonons induces no additional current we find from Eqs.(\ref{totcur},\ref{screen}) the screening factor Eq.(\ref{screening}). In this paper a connection between the divergence of the screening factor Eq.(\ref{screening}) at some critical power $W_c$ and the development of a zero resistance state is sought. Toward this end a realistic model of magneto-transport without (and then with) microwaves in GaAs heterostructures is required.

There are two well understood models of the magneto-transport of 2DEG in high Landau levels. First is the model with short range disorder and large scattering momenta: $ql_H\gg 1$, where $l_H$ is the magnetic length - the self-consistent Born approximation, exact in high Landau levels \cite{AFS}. It predicts the classical Drude conductivity tensor specified by the scattering time $\tau$ in weak magnetic fields $\omega_H\tau\ll 1$ and oscillating longitudinal conductivity with: $\sigma_{xx}^{\textrm{\small max}}(H)\sim e^2 N/\hbar$,\cite{kubo65} where $N$ is the Landau level index, in strong magnetic fields $\omega_H\tau\gg 1$. Second is the model with long range disorder and small scattering momenta: $qR_c\ll 1$.\cite{PG} It employs the approximate conservation of the adiabatic invariant and predicts the classical localization \cite{iotrug} with the negative magneto-conductivity: $\sigma_{xx}(H)\sim \exp(-H/H_c)$.\cite{fdps97} In GaAs heterostructures the disorder potential from Si donors has a cut-off in momentum space $q_c\sim 1/2d$ where $d$ is the spacer width. Thus, a crossover from the short range disorder at weaker magnetic fields to the long range disorder at stronger magnetic fields occurs at $l_H\sim 2d$ or $H_c\sim 0.3kG$ (for typical $d=80\textrm{nm}$). But above this $H_c$ experiments\cite{Mani} show positive rather than negative magneto-resistance. Second contradiction is the behaviour of the Shubnikov-de-Haas oscillations when the longitudinal resistivity approaches zero. The theory of long range disorder predicts the classical localization and wide platoes of quantum Hall zero resistivity whereas experiments \cite{zdpw,dsuk04} show no platoes but instead a cusp-like touch of zero resistivity value.

In this paper a model of magneto-transport in high mobility GaAs heterostructures with disorder that is a superposition of weak short range and strong long range potentials is introduced. The main result is that the local diffusion constant is {\it inversely} proportional to the local electric field $E$ of the long range potential. Indeed, this field lifts the degeneracy of the Landau level and impart a velocity and a linear dispersion to the electron: $\epsilon(p)= el_H^2\vec{E}\times \vec{p}$. Therefore the density of states is inverse to the electric field: $\nu(\epsilon_F)\sim 1/E$ (neglecting a logarithmic factor). The mean scattering rate $1/\tau$ is proportional to the mean square of the short-range potential $w$: $1/\tau=2\pi\nu(\epsilon_F)\langle w^2\rangle$, in analogy with the usual disordered metals. In magnetic field the diffusion constant is proportional to $1/\tau$ due to dominant Lorents force. 

In this paper the temperature is zero as it is appropriate for physics of disorder. It is assumed that in the state driven by microwaves the electron relaxation due to phonons is effective and keeps the electron distribution stationary, approximated by the Fermi-Dirac function. 

In section II a model of free electrons moving in disorder potential is introduced. In section III the diagrammatic method is described and the density of states and the conductivity tensor for three simplified disorder models are calculated. In section IV the magneto-transport of 2DEG in the realistic model with both long and short ranged disorder is found. In Section V the state driven by microwaves is considered and the divergent screening Eq.(\ref{screening}) is derived in the microscopic theory. It is explained how zero resistance states emerge.

\section{Model}

In the Fermi liquid theory the transport is described in the approximation of non-interacting quasiparticles. In strong magnetic fields and for long range potential this pictures requires explanation. In 2DEG with weakly broaden Landau levels in the long range disorder the chemical potential could not be a smooth function. Instead the Coulomb interaction makes the density a smooth function whereas the chemical potential is pinned by the Landau level and makes jumps along lines where the density crosses an integer filling values. It has been proven that the width of the quantum Hall stripes of integer filling is negligible in high Landau levels \cite{chshg92} (for $H=0.4kG$ and $n=3\times 10^{11}cm^{-2}$ the stripe width is $a=0.3 d\ll 2d$). In a superposition of long range and weak short range potentials a quasiparticle will diffuse inside a band around the energy level contour of the long range potential \cite{bla-bla}. The boundaries of this diffusion band are determined by the conservation of the total energy that is divided between the long range potential and a 'reservoir' of the short range energy (local Landau level broadening) \cite{bla-bla}. But interacting quasiparticles diffuse out of these band limits. One quasiparticle could climb up in the long range potential landscape whereas another quasiparticle falls down - very similar to the backflow in Fermi liquids. Therefore the magneto-transport in high Landau levels derives from an unrestricted quasiparticle diffusion. It destroys the classical localization and could be described in the non-interacting approximation. 

The Hamiltonian of free electrons in a random disorder potential $u(\vec{r})$ reads:
\begin{equation}\label{Hamiltonian2DEG}
\hat{H}=\int\psi^\dagger(\vec{r})\left[\frac{1}{2m}\left(i\hbar\vec{\nabla}+\frac{e}{c} \vec{A}(\vec{r}) \right)^2 +u(\vec{r})\right]\psi(\vec{r}) d^2\vec{r},
\end{equation}
where $m$ is the renormalized by interaction band mass, $\vec{A}(\vec{r})$ is the vector potential of the perpendicular magnetic field $H$. We use the Landau gauge $A_x=-Hy$ and suppress the spin index due to the vanishing exchange interaction in high Landau levels.

There are three physical parameters of 2DEG: $m$, $H$ and the density of electrons: $n=\langle \psi^\dagger(\vec{r}) \psi(\vec{r})\rangle= N/2\pi l_H^2$. One is the Fermi energy, whereas the two others are dimensionless: the Landau level number $N$ and the interaction parameter: $r_s=e^2m/\sqrt{\pi n}$. In spirit of Landau theory of Fermi liquids the model (\ref{Hamiltonian2DEG}) describes quasiparticles near the Fermi level. We use the magnetic units: $\hbar=1$, $e=c$, $H=1$, $\omega_H=1/m$. In high Landau levels ($N\gg 1$) there are three distinct lengths: the cyclotron radius $R_c= \sqrt{2N} l_H$, the magnetic length $l_H=\sqrt{c\hbar/eH}=1$ and the inter-electron distance $1/\sqrt{\pi n}$. 

A random electrostatic potential in the quantum well is created by remote ionized Si donors with the positive charge $e$. Let donors be confined to a narrow 2D layer - $\delta$-doped layer - parallel to the quantum well and separated by the spacer of width $d$ free from impurities. The positions of donors assumed to be uncorrelated and their areal density is equal to the density of electrons $n^{2D}_{\text{imp}}=n$. The correlation function of the donor electrostatic potential $u_0(\vec{r})$ reads:
\begin{equation}\label{SpacerIm}
S_0(\vec{q})=\langle u_0(\vec{q}) u_0(-\vec{q})\rangle=\left(\frac{2\pi e^2}{|\vec{q}|}\right)^2 n \exp\left(-2|\vec{q}|d\right),
\end{equation}
in the momentum range $q\ll\sqrt{n}$. We assume $d\sqrt{n}\gg r_s$. The donor random potential $u(\vec{r})$ is Gaussian with the probability distribution functional:
\begin{equation}\label{PDD}
\mathcal{P}[u]=\frac{1}{Z}\exp( -\sum_{\vec{q}} u(\vec{q})u(-\vec{q})/2S(q) ),
\end{equation}
where $Z$ is the normalization factor.

Electrons screen the donor potential $u_0(\vec{r})$ near the 2DEG plane according to Eq.(\ref{screen}). In high Landau levels a linear response occurs even in the quantum limit of non-overlapping Landau levels. The correlation function of electron density in response to $u_0(\vec{r})$ is: $\langle n(\vec{q}) n(-\vec{q}) \rangle =n \exp\left(-2|\vec{q}|d\right)$. The average density variation: $\langle (\delta n)^2 \rangle = n/2\pi (2d)^2$. For typical densities $n$ in GaAs heterostructures the screening is complete:
\begin{equation}\label{CorVV}
S(\vec{q}) =\langle u(\vec{q}) u(-\vec{q})\rangle =n \exp\left(-2|\vec{q}|d\right)/ \nu^2(\epsilon_F).
\end{equation}
This result is quasi-classical in nature but holds for high locally non-overlapping Landau levels as well.

Usually the areal density of Si in $\delta$-doped layer is much larger than $n^{2D}_{\text{imp}}$. Then ionized donors can adjust positions to lower the total electrostatic energy. Resulting configuration of ionized donors is different from the uncorrelated one and is better described by highly correlated liquid state. The correlation function is modified:
\begin{equation}\label{SpacerLiquidIm}
S(\vec{q})=n S_f(q) \exp\left(-2|\vec{q}|d\right)/\nu^2(\epsilon_F),
\end{equation}
where $S_f(q)$ is the structural function of the liquid state. Generally, $S_f(q)=A\vec{q}^2$ at small $\vec{q}$. We note that for periodic structures of ionized donors $S(\vec{q})=0$ for all $\vec{q}$ except on the Brave lattice points.

We expand the electron Green function into perturbative series of the disorder potential $u(\vec{r})$, using a diagrammatic method \cite{AGD}. We then average this series over the probability distribution density Eq.(\ref{PDD}). In the result we get diagrams with electron and impurity lines. The electron density vertex describes an electron scattering off the potential $u(\vec{q})$ from the state $(n,p)$ into a state $(n',p')$, where the indices $p$ and $p'$ counts degeneracy inside the Landau level $n$. The vertex is the product of the reduced vertex and the magnetic phase factor:
\begin{equation}\label{ImurityVertex}
V(np,n'p',\vec{q})=V_{nn'}(\vec{q})U(p,\vec{q})\delta_{p',p+q_y}.
\end{equation}
The magnetic phase factor is universal for all Landau level: $U(p,\vec{q})= \exp(iq_x (p+q_y/2))$,  whereas the reduced vertex depends on the incoming and outgoing Landau level indices and the scattering momentum:
\begin{equation}\label{VertexLaguerre}
V_{nn'}=\left(\frac{n'!}{n!}\right)^{\pm 1/2}\!\!\!\! \left(\frac{q_y\pm iq_x}{\sqrt{2}}\right)^{|n-n'|}\!\!\!\! L^{|n-n'|}_{\textrm{\small min}(nn')} \left(\frac{q^2}{2}\right)e^{\displaystyle -\frac{q^2}{4}},
\end{equation}
where $L_m^k(x)$ is the Laguerre polynomial, and where the sign $\pm$ in (\ref{VertexLaguerre}) corresponds to the cases: $n>n'$ and $n<n'$. In the limit $n,n'\rightarrow\infty$ the quasi-classical approximation applies provided $q\leq 4\sqrt{\pi n}$: 
\begin{equation}\label{VertexBessel}
V_{nn'}(\vec{q})=\left(\frac{q_x+iq_y}{|\vec{q}|}\right)^{n-n'} \sqrt{\frac{\phi_{nn'}(q)}{qp_{nn'}(q)}} J_{|n-n'|}\left(\phi_{nn'}(q)\right),
\end{equation}
where $p_{nn'}(q)=\sqrt{n+n'+1-q^2/4}$ is the quasi-classical momentum and $\phi_{nn'}(q)= qp_{nn'}(q)/2+(n+n'+1)$ $\arcsin(q/2\sqrt{n+n'+1})$ is the quasi-classical phase. 
 
The electron propagation in static potential is either retarded or advanced. The bare Green function is diagonal in the Landau level index:
\begin{equation}\label{GreenBare}
g_n^{R,A}(\epsilon)=1/(\epsilon-n\omega_H\pm i\delta),
\end{equation}
where $0<\delta\rightarrow 0$ is the bare dissipation. In the Schwinger proper time representation:
\begin{equation}\label{GreenSchwinger}
g^R_n(\epsilon)=-i\int_0^\infty\exp\left(i(\epsilon-n\omega_H)t-\delta t\right)\ dt.
\end{equation}
The advanced Green function is the complex conjugate of the retarded one: $g^A_n(\epsilon)=\left(g^R_n(\epsilon)\right)^*$.

We include two electron density vertices into the impurity line that connects them. Thus the impurity line that is not crossed by other impurity lines reads: 
\begin{equation} \label{ImpurityLine}
u_{k}(N)=\int S(\vec{q})\left|V_{N,\ N+k}(\vec{q})\right|^2\ \frac{d^2\vec{q}}{(2\pi)^2}.
\end{equation}
In this notation the Green functions and the impurity lines does not depend on the degeneracy index $p$. 

If the two impurity lines with wavevectors $\vec{q}_1$ and $\vec{q}_2$ intersect each other then the total integrand given by the product of the two integrands in Eq.(\ref{ImpurityLine}) is entangled by a complex magnetic phase factor: $\exp(i\vec{q}_1\times \vec{q}_2)$. This magnetic phase factor can be conveniently represented as a commutator:
\begin{equation}\label{MagPhaseCom}
\exp\left(i\vec{q}_1\times \vec{q}_2\right)=\left[\hat{V}_M(\vec{q}_1)\ \hat{V}_M(\vec{q}_2) \right]_- ,
\end{equation}
of the two magnetic phase operators:
\begin{equation}
\hat{V}_M(\vec{q})=\exp\left(i(q\hat{a}^\dagger+ q^*\hat{a})/\sqrt{2}\right),
\end{equation}
that substitutes the magnetic phase in Eq.(\ref{ImurityVertex}) and is expressed in terms of the Bose operators $\hat{a}^\dagger$ and $\hat{a}$. This operator acts in the space of the Landau level degeneracy index $p$ in the Landau gauge. 

\section{Density of states and conductivity}

In this section we consider three artificial models of disorder using two cutoff wavevectors: $q_L\ll 1/l_H$ and $q_S\gg 1/l_H$. A short range potential model S is the model (\ref{Hamiltonian2DEG}) with the correlation function $S(q)=0$ for $q<q_S$. A long range potential model L is the model (\ref{Hamiltonian2DEG}) with the correlation function $S(q)=0$ for $q>q_L$. A critical model C is the model (\ref{Hamiltonian2DEG}) with the special correlation function: $S(q)= A/q$, for all wavevectors from $1/R_c$ to $\sqrt{\pi n}$. In high Landau levels around $N$ the correlation function $S(q)$ is represented by the set of coefficients $u_k(N)$ Eq.(\ref{ImpurityLine}). We expand the average retarded Green matrix $G^R_{nn'}(\epsilon)$ into a series of coefficients $u_k$, using a diagrammatic method \cite{AGD}. The density of states is related to the imaginary part of this Green function \cite{AGD}:
\begin{equation}\label{DOSImage}
\rho(\epsilon)=-\text{Im} \text{Tr}\langle G^R(\epsilon)\rangle /\pi.
\end{equation} 

\textbf{Density of states} for non-overlapping Landau levels in the limit $\sqrt{u_0}/\omega_H\ll 1$ and $u_k\ll u_0$ for $k>0$. Accordingly we set $u_k=0$ for $k\geq 1$ and $u_0=u$. Thus the Green function is diagonal in the Landau level index $N$ and we omit it from the general expansion:
\begin{equation}\label{GreenExpantion}
\langle G^R(\epsilon)\rangle =g^R(\epsilon)\sum_{k=0}^{\infty}c_k u^k \left(g^R(\epsilon) \right)^{2k},
\end{equation}
where the combinatorial coefficients $c_k$, with $c_0=1$, are determined by the magnetic phase factors and  $g^R(\epsilon)$ is the bare Green function (\ref{GreenBare}). In the limit $N\rightarrow\infty$ there are just three such unique series $c_k$ corresponding to the three different models: C, L, S. 

In model S the magnetic phase factor for diagrams with crossed impurity lines is small as $1/N$. The self-consistent Born approximation neglects crossing and coefficients $c_k$ are the Catalan numbers (1,2,5,14,42...) and the averaged Green function reads \cite{AFS}:
\begin{equation}\label{GreenS}
\langle G^R(\epsilon)\rangle_S = -\frac{1}{2ug(\epsilon)}-\frac{i}{u} \sqrt{u-\frac{1} {4g^2(\epsilon)}},
\end{equation}
where $g^{-1}(\epsilon)=\epsilon-N\omega_H$.

In model L the the magnetic phase for crossing of two impurities lines is zero. Therefore the coefficient $c_k$ is given by the combinatorial number of all pairing of $2k$ points: $c_k=(2k-1)!!$. The average Green function (\ref{GreenExpantion}) has been summed up in the Ref.\cite{rs93}:
\begin{equation}\label{GreenL}
\langle G^R(\epsilon)\rangle_L=-i\int_0^\infty\exp\left(-\frac{ut^2}{2}+ i\frac{t}{g(\epsilon)}\right)dt. \end{equation}

For critical model C each integral over impurity line wavevector gives a logarithmic factor $\log(R_c\sqrt{\pi n})$. In the leading logarithm approximation we have calculated the nine first coefficients $c_k$ and they coincide exactly with that of the Wegner model for 'delta'-correlated impurity potential on the lowest Landau level \cite{Wegner}:
\begin{equation}\label{GreenC}
\langle G^R(\epsilon)\rangle_C=\frac{\partial}{\partial\epsilon} \log\int_0^\infty\exp \left( -\frac{ut^2}{4}+ i\frac{t}{g(\epsilon)}\right)dt,
\end{equation}
despite the fact that the model C is defined in high Landau levels $N\gg 1$.

The density of states Eq.(\ref{DOSImage}) is the semi-circle law in the case of the model S, the Gaussian nonuniform broadening in the case of the model L \cite{rs93}, and the Wegner distribution \cite{Wegner} for the critical model C. 

\textbf{Arbitrary Landau level mixing.} For long range model L we prove Eq.(\ref{GreenL}) in the case of arbitrary Landau level mixing. This subsection essentially rederives results of Ref.\cite{rs93} using an alternative method. For model L the electron Green function can be represented in the Schwinger proper time as: 
\begin{equation}\label{GreenLSchwinger}
G^R_{nn'}=-i\int\!\!\int_0^\infty\!\! dt\ U_{nn'}(t)\ e^{\displaystyle i\epsilon t-\frac{ut^2}{2}-\sum_{k>1} \frac{|\psi_k|^2}{2}} \frac{d^2\psi_k}{2\pi},
\end{equation}
where $u=u_0$, $k>1$ and the evolution matrix:
\begin{equation}\label{Evolution}
U_{nn'}(t)=\langle n|\exp(-i\hat{h}t)|n'\rangle,
\end{equation}
is the unitary evolution operator of the Hamiltonian
\begin{equation}\label{hamLLM}
\hat{h}=-i\omega_H\frac{\partial}{\partial \phi}+\sum_{k=1}^\infty \sqrt{u_k}\left(\psi_k e^{i\phi}+ \psi^*_k e^{-i\phi}\right),
\end{equation}
where $\psi_k$ are complex parameters. This Hamiltonian accounts for the Landau level mixing at arbitrary values of $u_k/\omega_H$. The first term in (\ref{hamLLM}) is the kinetic energy whereas the non-diagonal second term accounts for scattering transfer between Landau levels. The convergent integrals over complex variables $\psi_k$ ($|\psi_k|\sim 1$) give a combinatorial number of crossed line pairing and the term $ut^2/2$ generates scattering processes that conserve the Landau level index. The wave function
\begin{equation}
\chi_j(\phi)=\frac{1}{Z}\exp\left(\sum_{k=1}^\infty\frac{\sqrt{u_k}}{ik}\left(\psi_k e^{i\phi}- \psi^*_k e^{-i\phi}\right)+i\omega_Hj\right)
\end{equation}
where $Z$ is the normalization factor is the solution of the Schroedinger equation $\hat{h} \chi(\phi)= \epsilon\chi(\phi)$ with eigenvalue $\epsilon=j\omega_H$. The condition that the wave function is single valued restricts $j$ to be an integer. 

We prove that eigenvalues of Hamiltonian (\ref{hamLLM}) are $j\omega_H$ exactly. Consider  shift operators: $\hat{s}_k= \exp(ik\phi)$. They commute with the Hamiltonian as: $[\hat{h},\hat{s}_k]= k\omega_H \hat{s}_k$. Let $\chi$ be the eigenfunction of $\hat{h}$ with the eigenvalue $\varepsilon$ then the function $\hat{s}_k\chi$ shifted by $k$ Landau levels is also the eigenfunction of $\hat{h}$ with the eigenvalue $\varepsilon+k\omega_H$. These functions are orthogonal: $\langle\chi^*\hat{s}_k\chi\rangle=0$ for $k\neq 0$. Consider the matrix of size $2N+1$ reducing the Hamiltonian (\ref{hamLLM}) in the basis $\exp(in\phi)$, where $n=-N..N$. Apart from few boundary eigenvalues all middle eigenvalues are equidistant. The trace of the Hamiltonian (\ref{hamLLM}) in this basis is obviously zero. The energy shift of the boundary eigenvalues is limited provided all $|\psi_k|\sim 1$. Therefore the energy shift of middle eigenvalues is bounded by $1/N$. 

Let $\chi_j(n)$ be the eigenfunction of the Hamiltonian (\ref{hamLLM}) corresponding to the middle eigenvalue $j\omega_H$ in the basis of functions $|n\rangle=\exp(in\phi)$. These eigen functions can be chosen to be real. Then we rewrite the Eq.(\ref{GreenLSchwinger}) as
\begin{equation}\label{GreenDrift}
G^R_{nn'}(\epsilon) =-i\int_0^\infty\sum_j \chi_j(n)\chi_j(n') e^{\displaystyle i(\epsilon-j\omega_H) t-\frac{ut^2}{2}} dt.
\end{equation}
Using the orthogonal $\sum_j \chi_j(n)\chi_j(n')=\delta_{nn'}$ and normalization $\langle\chi_j| \chi_j\rangle=1$ conditions we find that the density of state for Eq.(\ref{GreenDrift}) is given by Eq.(\ref{GreenL}).

\textbf{Conductivity in model S.} The conductivity tensor in the model S has been calculated in both cases of the large \cite{kubo65} and small \cite{ando74} Landau level mixing. The Drude conductivity depends on the transport scattering time $\tau_{\text{tr}}$: $\sigma_{aa}=e^2n\tau_{\text{tr}}/m$, where $aa$ is either $xx$ or $yy$. Delta correlated disorder potential $S(q)=2\pi u$ is a special one with the quantum scattering time $\tau=\tau_{tr}$. But in general $\tau<\tau_{tr}$. In the absence of magnetic field the usual ladder diagrams account for the difference between $\tau$ and $\tau_{tr}$. In relevant magnetic fields: $\sigma_{aa}\ll \sigma_{xy}$, the Drude formula reads $\sigma_{aa}=e^2n/ \omega_H^2\tau_{tr}$. The diagram in Fig.1a gives the conductivity $e^2n/ \omega_H^2\tau$ whereas the diagram in Fig1.b gives the transport correction. For $\delta$-correlated impurities the diagram Fig.1b is zero whereas for long range disorder the two diagrams Fig.1 cancels each other in the leading order.

\begin{figure}
\includegraphics{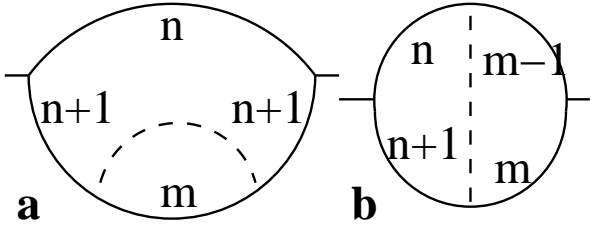}
\caption{Two diagrams: a and b, that together give the longitudinal conductivity in the limit of vanishing Landau level mixing. Dashed line is the short range potential impurity line. Full lines are electron propagations with the Landau level index being indicated.}
\end{figure}

\textbf{Conductivity in model L}. We use the Keldysh method to find that the longitudinal conductivity is explicitly a real function: $\sigma_{xx}(\Omega)= \sigma(-\Omega)^*$. We separate it into the reactive and the inductive parts: $Z(\Omega)= R(\Omega^2)+i\Omega L(\Omega^2)$, and we will not consider the inductive term here. For the reactive part we find:
\begin{equation}\label{SigmaKG}
\sigma_{aa}(\Omega)=\frac{e^2\omega_H^2}{\pi}\int_{-\infty}^{\infty}\!\frac{d\epsilon}{2\pi}\  \frac{f(\epsilon)-f(\epsilon-\Omega)}{\Omega} \sigma_{aa}(\Omega,\epsilon),
\end{equation}
where $f(\epsilon)$ is the occupation number of states with energy $\epsilon$. The effective conductivity of this state reads:
\begin{eqnarray}\label{SigmaXX}
\sigma_{aa}(\Omega,\epsilon)= \sum_{nn'=1}^\infty\! \sqrt{n'n}\ \ \langle \textrm{Im} G_{nn'}(\epsilon-\Omega)\nonumber\\ \textrm{Im} G_{n-1,n'-1}(\epsilon)+\textrm{Im} G_{nn'}(\epsilon) \textrm{Im} G_{n-1,n'-1}(\epsilon-\Omega)  \rangle ,
\end{eqnarray}
where the imaginary part of the Green function is: $\textrm{Im} G_{nn'}(\epsilon)= i(G^R_{nn'}(\epsilon)- G^A_{nn'}(\epsilon))/2$. We use the diagrammatic method \cite{AGD} to expand the disorder average in the effective conductivity Eq.(\ref{SigmaXX}) into a series of $u_k$. 
\begin{eqnarray}\label{Sigep}
\sigma^{xx}(\Omega,\epsilon)=\int\! \int_{-\infty}^\infty\int_{-\infty}^\infty \textrm{Re} \textrm{Tr} \left( \hat{U}(t)\hat{\Pi}^\dagger  \hat{U}(s)\hat{\Pi}\right) \nonumber\\  e^{\displaystyle i\epsilon t+i(\epsilon-\Omega)s -\frac{u(t+s)^2}{2}-\sum_{k} \frac{|\psi_k|^2}{2}}dt ds\ \frac{d^2\psi_k}{2\pi},
\end{eqnarray}
where $\hat{U}(t)=\exp(-i\hat{h}t)$ is the unitary evolution operator of the Hamiltonian (\ref{hamLLM}). The operator $\hat{\Pi}$ is the raising Landau level operator with the matrix elements $\Pi_{n,n-1}= \sqrt{n}$ and $\hat{\Pi}^\dagger$ is the Hermitian conjugated operator. The integrals with respect to $t,s$ from $-\infty$ to $0$ give advanced Green functions and the integrals from $0$ to $\infty$ give the retarded Green functions. Next we make use of the fact that $\chi_j(n)$ is the eigenfunction of the Hamiltonian (\ref{hamLLM}) with the eigenvalue $j\omega_H$. We rewrite Eq.(\ref{Sigep}):
\begin{eqnarray}\label{Sigep1}
\sigma^{xx}(\Omega,\epsilon)= \sum_{j,k}\langle \chi_j\hat{\Pi}^\dagger\chi_k\rangle \langle \chi_j\hat{\Pi}\chi_k\rangle \int_{-\infty}^\infty\int_{-\infty}^\infty e^{-\delta(|t|+|s|)}  \nonumber\\  e^{\displaystyle i(\epsilon-j\omega_H) t+i(\epsilon-\Omega-k\omega_H)s -\frac{u(t+s)^2}{2}}dt ds,
\end{eqnarray}
where the bare dissipation term with $0<\delta\rightarrow 0$ is added for the integral convergence. The sum over $j$ and $k$ is restricted: $j=k\pm 1$. The integral over $t+s$ gives the density of states whereas the integral over $s$ gives the Dirac delta functions: $\delta(\Omega\pm\omega_H)$. Therefore if the magnetic phase is neglected then only one of the two Green functions in the Kubo-Greenwood formula for conductivity: $\sigma= e^2 \hat{v}\textrm{Im} G \hat{v}\textrm{Im} G$ is broaden. Equivalently the diffusion constant in the Einstein formula: $\sigma(\Omega)= e^2\nu(\epsilon_F)D(\Omega)$, is that of the ideal electrons: $D(\Omega)\sim \delta(\Omega\pm\omega_H)$. To account for the magnetic phase in the model L would probably require a non-perturbative approach.

In the same way we find the Hall conductivity:
\begin{equation}\label{SigmaHall}
\sigma_{xy}(\Omega)=\frac{e^2\omega_H^2}{\pi}\int_{-\infty}^{\infty}\frac{d\epsilon}{2\pi} \ f(\epsilon)\frac{\sigma^{xy}(\epsilon,\Omega)-\sigma^{xy}(\epsilon,-\Omega)}{\Omega},
\end{equation}
where the Hall conductivity of the electron state $\epsilon$ is:
\begin{eqnarray}\label{Sigmaxy}
\sigma^{xy}(\epsilon,\Omega)=\sum_{nn'=1}^\infty\!\sqrt{n'n}\ \ \langle \textrm{Re} G_{n-1,n'-1}(\epsilon+\Omega) \nonumber\\ \textrm{Im} G_{nn'}(\epsilon) - \textrm{Re} G_{nn'}(\epsilon+\Omega) \textrm{Im} G_{n-1,n'-1}(\epsilon) \rangle,
\end{eqnarray}
where the real part of the Green function: $\textrm{Re} G_{nn'}(\epsilon)= (G^R_{nn'}(\epsilon)+ G^A_{nn'}(\epsilon))/2$. We expand the impurity average in Eq.(\ref{Sigmaxy}) into a series over coefficients $u_k$ and the result is partially summed up as:
\begin{eqnarray}\label{SiHep}
\sigma^{xy}(\epsilon,\Omega)=\int\! \int_{-\infty}^\infty\!\int_{-\infty}^\infty \textrm{Im} \textrm{Tr} \left( \hat{U}(t)\hat{\Pi}^\dagger  \hat{U}(s)\hat{\Pi}\right)\textrm{sgn}(s) \nonumber\\ e^{\displaystyle i\epsilon t+i(\epsilon+\Omega)s -\frac{u(t+s)^2}{2}-\sum_k \frac{|\psi_k|^2}{2}}dt ds\frac{d^2\psi_k}{2\pi}.
\end{eqnarray}
In terms of the eigenfunction of the operator $\hat{U}(t)$ (\ref{Evolution}):
\begin{eqnarray}\label{SiHep1}
\sigma^{xy}(\epsilon,\Omega)= \sum_{j,k} \left(\langle\chi_j \hat{\Pi}^\dagger\chi_k\rangle\langle\chi_k \hat{\Pi}\chi_j\rangle -c.c. \right)\int_{-\infty}^\infty dt ds \nonumber\\  \textrm{sgn}(s)\  e^{\displaystyle i(\epsilon-j\omega_H) t+i(\epsilon+\Omega-k\omega_H)s -\frac{u(t+s)^2}{2}} .
\end{eqnarray}
Now we make use of the matrix element: $\langle\chi_j\hat{\Pi}^\dagger \chi_k\rangle= \sqrt{j}\delta_{j,k+1}$, where $\delta_{j,k}$ is the Kroneker symbol. Taking the integrals over $t$ and then $s$ we find: $\sigma^{xy} (\epsilon,\Omega)=\pi \rho(\epsilon)/ (\omega_H+\Omega)$, where the density of states:
\begin{equation}\label{DOSL}
\rho(\epsilon)=\frac{1}{\sqrt{2\pi u}} \sum_j\exp\left(- \frac{(\epsilon-j\omega_H)^2}{2u}\right).
\end{equation} 
The Hall conductivity is given by the Eq.(\ref{SigmaHall}):
\begin{equation}\label{Hall}
\sigma_{xy}(\Omega)=\frac{e^2\omega_H^2}{\omega_H^2+\Omega^2}\int_{-\infty}^\infty \frac{d\epsilon}{2\pi}\ f(\epsilon)\rho(\epsilon).
\end{equation}
The integral over $\epsilon$ in Eq.(\ref{Hall}) gives the total number of filled Landau levels: $N=2\pi l_H^2 n$. At $\Omega=0$ we find $\sigma_{xy}=\sigma^H=ecn/H$. This result explicitly does not depend on the electron distribution function $f(\epsilon)$. 

\section{Two lengths disorder model}

Scattering rate off impurities is proportional to the scattering cross-section and the velocity of incoming particle. The velocity of 2D electron in magnetic field is zero. It is disorder that creates the dispersion $\epsilon(p)$ - Landau level broadening - and imparts random velocity to electrons: $v=d\epsilon(p)/dp$. In the previous Section the long range disorder was shown to be ineffective in broadening the Landau level locally. Thus we add a weak short range disorder to overcome this problem. No microwave excitation is considered in this Section.

The origin of short range disorder in GaAs heterostructures may lie in the barrier layer $Ga_{1-x}Al_x As$. Because Al atoms are distributed randomly the local energy barrier fluctuates as $\sim 1/\sqrt{N_{im}}$ where $N_{im}$ is the typical number of Al atoms in the volume of typical electron wave function $100A$. The short range disorder potential $w(\vec{r})$ is assumed Gaussian and to have the correlation function: $S_w=\langle w(\vec{q}) w(-\vec{q})\rangle= 1/2\pi\nu(\epsilon_F)\tau_w= 1/m\tau_w$. Using the vertex Eq.(\ref{VertexLaguerre}), the definition Eq.(\ref{ImpurityLine}) and the orthogonal and normalization properties for Laguerre polynomes we find the 'short range' impurity line in $N$-th Landau level: $w=w_{NN}=S_w/2\pi l_H^2=\omega_H/2\pi \tau_w$. We assume that $w\ll u$, but $w$ could be comparable to $u_{tr}$.

In the limit $l_H\ll 2d\ll R_c$ the conductivity of the two lengths disorder model could be found. We neglect the diagrams with crossed short-short and short-long impurity lines whereas the magnetic phase for crossed long-long impurity lines is assumed zero. The average Green function is determined by the equation:
\begin{equation}\label{TwoGreen}
G^R(\epsilon)=-i\int_0^\infty \exp\left(-\frac{u}{2}t^2+i\epsilon t-i\Sigma^R(\epsilon)t\right) dt,
\end{equation}
where $\Sigma^R(\epsilon)=wG^R(\epsilon)$ is the self-energy of electron in local long-range potential before averaging. Solution of Eq.(\ref{TwoGreen}) satisfies the normalization condition for any $w$: $\int_{-\infty}^{\infty}\textrm{Im} G^A(\epsilon)\ d\epsilon=\pi$

The longitudinal conductivity is given by diagrams in Fig.1 where the impurity line is necessarily the short range line. Remember that the unitary rotation $\hat{U}(t)$ Eq.(\ref{Evolution}) eliminates for long range potential all transitions between Landau levels such as in Fig.1. The diagram Fig.1b with the short range line is zero. Using unitary rotation Eq.(\ref{Evolution}) we simplify the two current vertices and the Green functions in the diagram Fig.1a and find the longitudinal conductivity:  
\begin{equation}\label{twocond}
\sigma_{aa}=e^2\frac{2N}{\pi^2} w\ \langle\textrm{Im} G^R(\epsilon_F)\rangle\langle\textrm{Im} G^R(\epsilon_F)\rangle.
\end{equation}
This equation coincides with that in the phenomenological force-force correlation approach \cite{ll03,lt85}, but we derived the limits of applicability of Eq.(\ref{twocond}). In the quantum limit of non-overlapping Landau levels $\omega_H\gg \sqrt{u}$ at higher magnetic fields: 
\begin{equation}\label{quantumtwocond}
\rho_{aa}=\frac{m}{e^2n\tau_w}\frac{\omega_H^2}{2\pi u}\ e^{\displaystyle -\frac{(\epsilon_F- N\omega_H)^2}{u}}
\end{equation}
whereas in the classical limit $\omega_H\ll \sqrt{u}$ in weaker magnetic fields we find the Drude formula: 
\begin{equation}\label{classtwocond}
\rho_{aa}=m/e^2n\tau_w.
\end{equation}

Corrections to Eqs.(\ref{quantumtwocond},\ref{classtwocond}) could arise due to a possible additional contribution to the long range disorder originating from very long range technological imperfections at $qR_c\ll 1$. Accordingly we distinct the technological and the donor disorders: $u=u_\infty +u_{2d}$. In the case $u_\infty\gg u_{2d}$ there will be a second crossover at $\omega_H\sim \sqrt{u_{2d}}$ from the classical Drude regime Eq.(\ref{classtwocond}) at weaker magnetic fields to a classical positive magneto-resistivity:
\begin{equation}\label{positiveMR}
\rho_{aa}=\frac{m}{e^2n\tau_w}\ \frac{\omega_H}{\sqrt{2\pi u_{2d}}}
\end{equation}
at higher magnetic fields: $\sqrt{u_{2d}}\ll \omega_H\ll \sqrt{u_\infty}$. Because $l_H\ll 2d\ll R_c$, $\sqrt{u_{2d}}\sim\sqrt{H}$ and, therefore, $\rho_{xx}\sim \sqrt{H}$ - such positive magneto-resistivity behaviour is often seen on experimental traces.

\textbf{Drifting states.} For long range potential $2d\gg R_c$ states of electron transform into drifting states along the energy levels and during a short time locally the long range potential could be approximated by uniform electric field crossed with magnetic field. Crossing of 'long and short range' impurity lines now has magnetic phase close to zero and one has to account for multiple crossings to build up the magnetic phase. Formally this can be described using the multi-field Hubbard -Stratanovich decomposition of the action averaged over the long range potential that includes the impurity line weights and the magnetic phase factors:
\begin{eqnarray}\label{AverAction}
S[\psi]=-\frac{1}{2}\int dt dt' \sum_{\vec{q},n_1n_2n_3n_4} S(\vec{q}) V_{n_1n_2}(-\vec{q}) V_{n_3n_4}(\vec{q}) \nonumber\\ \psi_{n_1}^\dagger(t) \hat{V}_M(-\vec{q}) \psi_{n_2}(t) \ \psi_{n_3}^\dagger(t') \hat{V}_M(\vec{q}) \psi_{n_4}(t'),
\end{eqnarray}
where the time $t$ is the Schwinger proper time in Eq.(\ref{GreenL}) and operators $\hat{V}_M$ act on the Landau level degeneracy index. For simplicity scattering that does not change the Landau level number $N$ is considered. The momentum integral in the action is performed symbolically:
\begin{eqnarray}\label{SymblAction}
S[\psi]/\mathcal{A}= -\frac{1}{2}u\int\int dt dt'\ \psi^\dagger(t)\psi^\dagger(t') \nonumber\\ :\left(1+ \frac{l_H^2}{2d^2}(\hat{a}^\dagger-\hat{b}^\dagger) (\hat{a}-\hat{b})\right)^{-1/2}: \psi(t')\psi(t),
\end{eqnarray}
where $\mathcal{A}$ is the area of system, $:\ ...\ :$ stands for normal ordering and operators $\hat{a},\hat{b}$ act at times $t,t'$ correspondingly. Next we use the condition $qR_c\ll 1$ and assumption of short time electron propagation, with the expectation values of $\hat{a}$ and $\hat{a}^\dagger$ being much smaller then $2d/l_H$. Therefore we expand the action (\ref{SymblAction}):
\begin{equation}
S[\psi]/\mathcal{A}= -\frac{1}{2} u\sum_{i,j=0}^{\infty} \left(\frac{l_H}{2d}\right)^{2i+2j}\ \hat{O}_{ij}(t) \hat{O}_{ji}(t')
\end{equation}
where for short times the operator $\hat{O}_{mn}$ has the asymptote: $\hat{O}_{mn}(t)= C_{nm}\psi^\dagger(t) \left(\hat{a}^\dagger\right)^m \left(\hat{a}\right)^n \psi (t)$, where $C_{nm}$ are combinatorial coefficients. Next we make the Hubbard -Stratanovich transformation using fields $\phi_0$, $\phi_1$ ...: 
\begin{equation}\label{MFlong}
S[\psi]=-i\int \hat{H}_{lr}\  dt \ -\phi_0^2/2-|\phi_1|^2-...,
\end{equation}
where $\mathcal{A}=1$ and the effective long range Hamiltonian is:
\begin{equation}\label{HamLRMF}
\hat{H}_{lr}=\sqrt{u}\psi^\dagger\left(\phi_0  +\frac{l_H}{2d}\frac{1}{\sqrt{2}} \left(\phi_1\hat{a}^\dagger+ \phi^*_1\hat{a} \right)+...\right) \psi .
\end{equation}
Note that Eq.(\ref{MFlong}) is an analog of the mean-field (saddle) approximation in the usual impurity scattering theory.

\begin{figure}
\includegraphics{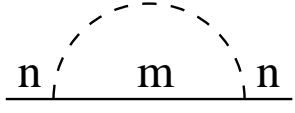}
\caption{The self-energy. Dashed line is the short range potential impurity line. Full lines are electron propagation.}
\end{figure}

Next we find the Green function averaged over the short range disorder. 'Short range' impurity lines do not cross due to the vanishing magnetic phase factor and the self-energy is exactly one diagram in Fig.2 averaged over the long range disorder eventually. In the mean-field approximation for the long range disorder Eq.(\ref{HamLRMF}) the magnetic phase factors of the two vertices in Fig.2 are entangled with the electron propagation: $\langle 0| \exp(i(Q^*\hat{a}+Q\hat{a}^\dagger)/\sqrt{2}) \exp(it\hat{H}_{lr}) \exp(-i(Q^*\hat{a}+Q\hat{a}^\dagger)/\sqrt{2}) \exp(-it\hat{H}_{lr}) |0\rangle $. Calculating this factor we find the self-energy:
\begin{equation}
\Sigma(t)=\int wV^2_{NN}(\vec{Q})\ e^{\displaystyle i\frac{\sqrt{u}t}{4d} (\phi_1 Q^*+\phi_1^* Q)} \frac{d^2\vec{Q}}{2\pi}.
\end{equation}
Using the vertex in terms of Laguerre polynome Eq.(\ref{VertexLaguerre}) we find: $\Sigma(t)=w\left(L_{N} (x^2/2)\right)^2 \exp(-x^2/2)$, where $x=\sqrt{u}|\phi_1|t/2d$. In high Landau levels $N\gg 1$ important is the evolution time such that $\sqrt{u}t\ll 1$ and $\sqrt{2N}x\gg 1$. Thus we use the large argument asymptote for Laguerre polynome: $\Sigma(t)=wJ_0^2(\sqrt{2N}x)$, and Bessel function (neglecting fast oscillations):
\begin{equation}
\Sigma(t)=\frac{\Delta}{|t|},\ \ \ \Delta=\frac{w}{\pi\mathcal{E}}, \ \ \  \mathcal{E}=\frac{\sqrt{u}|\phi_1|R_c}{2d}.
\end{equation}
For retarded Green function $t>0$ and further solution of the Dyson equation proceeds exactly as in Ref.\cite{kash04}:
\begin{equation} \label{AvG}
\epsilon=\frac{1}{G^R(\epsilon,\phi_1)}-i\Delta\log\left(\frac{\mathcal{E}}{\Delta- iG^{-1}_R(\epsilon,\phi_1)}\right).
\end{equation}
In the limit $w\ll uR^2_c/(2d)^2$ we find $\textrm{Im}G^R(t,\phi_1)=\exp\left(-\Delta t\log(\mathcal{E}/\Delta)\right)$. The semi-circle density of states transforms into the Lorentz density of states. The average over long range potential Green function is:
\begin{equation}\label{avLRGreen}
\langle G^R(\epsilon)\rangle = \int \frac{d^2\phi_1}{\pi}\ G^R(\epsilon,\phi_1).
\end{equation}
This result could be derived in alternative way. In high Landau levels scattering for electron drifting in the long range random potential $u(\vec{r})$ reads:
\begin{equation}\label{AGtau}
\frac{1}{2\tau}=w\int \textrm{Im}G(\epsilon-Q_y \partial_y u) |V_{NN}(\vec{Q})|^2 \frac{d^2\vec{Q}}{(2\pi)^2},
\end{equation}
where we used Landau gauge with $y$ axis being along the long range electric field. Evaluating Eq.(\ref{AGtau}) we find:
\begin{equation}\label{AGtaures}
\frac{1}{\tau}=\frac{w}{\pi^2 R_c |\vec{\partial}u|}\log\left(\frac{R_c}{2d}\right).
\end{equation} 

\textbf{Diffusion between drifting states}. We use the Einstein relationship and calculate the diffusion constant from the diagrams in Fig.1. Averaging over very long range potential $qR_c\ll 1$ is done using the mean field approximation Eq.(\ref{MFlong}). We parameterize Green functions by the Schwinger proper times $t$ and $s$ and write down the conductivity for example for the diagram in Fig.1a (after the unitary rotation Eq.(\ref{Evolution}) that eliminates the Landau level mixing - $m=n$):
\begin{eqnarray}
\sigma_{Fig.1a}=\frac{e^2}{\pi}\int \frac{d\phi_0}{\sqrt{2\pi}}\frac{d^2\phi_1}{\pi}\ e^{\displaystyle -\frac{\phi_0^2}{2}-|\phi_1|^2} \int \frac{d^2\vec{Q}}{(2\pi)^2}S_w \nonumber\\
V^2_{NN}(\vec{Q})\ \langle 0|\ \textrm{Im}\int_0^\infty dt e^{\displaystyle -it\hat{H}_{lr}[a^\dagger,a]}\ e^{\displaystyle i\frac{Q^*\hat{a}+Q\hat{a}^\dagger}{\sqrt{2}}}\nonumber\\ \textrm{Im}\int_0^\infty ds\  e^{\displaystyle -is\hat{H}_{lr}[a^\dagger,a]}\ e^{\displaystyle -i\frac{Q^*\hat{a}+Q\hat{a}^\dagger}{\sqrt{2}}} |0\rangle .
\end{eqnarray} 
The integral with respect to $t+s$ is convergent in very short times $\sim 1/\sqrt{u}$ and gives the density of states that is finally omitted from the diffusion constant. Then we set $s=-t$ and calculate the matrix element in the space of boson operators $\hat{a}^\dagger \times\hat{a}$. It is identical as for the self-energy above. We do the same for the diagram in Fig.1b and sum up both. Finally we find the anisotropic diffusion constant with two components along and perpendicular to the local long range electric field ($\vec{\phi}_1$):
\begin{equation} \label{ConductivityLR}
D_{l,t}= Nw\int \frac{d^2\phi_1}{\pi}\int_0^\infty dt\ \left\{ \begin{array}{c} P(t) \\ S(t) \end{array} \right\} \textrm{Im} G^R(t,\phi_1)
\end{equation}
where $P(t)=J_0^2(\mathcal{E}t)- J_1^2(\mathcal{E}t)$ and $S(t)= J_0(\mathcal{E}t) J_2(\mathcal{E}t)+ J_1^2(\mathcal{E}t)$. Using Eq.(\ref{AvG}) we evaluate the 'diffusion' Green function as: $\textrm{Im}G^R(t)=\exp\left(-\Delta t\log(\mathcal{E}/\Delta)\right)$. Diffusion constants for a given local long range field $\vec{\phi}_1$ are:
\begin{eqnarray}
D_l(\vec{\phi}_1)=\frac{2Nw}{\pi \mathcal{E}}\left(\frac{1}{k}E(k) -\frac{\Delta_L^2k}{4\mathcal{E}^2} K(k)\right) \nonumber\\
D_t(\vec{\phi}_1)=\frac{2Nw}{\pi \mathcal{E}}\left(\frac{1}{k}E(k)+ \frac{\Delta_L^2k}{4\mathcal{E}^2}K(k) -\frac{\pi\Delta_L}{2\mathcal{E}}\right),
\end{eqnarray}
where $K(k)$ and $E(k)$ are Jacobi elliptic functions with modulus $k= 2\mathcal{E}/ \sqrt{\Delta_L^2+4\mathcal{E}^2}$ and $\Delta_L= \Delta\log(\mathcal{E}/\Delta)$. In the limit of weak short range disorder: $w\ll uR^2_c/(2d)^2$, the diffusion constant becomes isotropic:
\begin{equation}\label{DiffD}
D=\frac{2dR_cw}{\sqrt{\pi}\sqrt{u}}.
\end{equation}
Thus we derived that for very long range disorder the diffusion constant is inversely proportional to the average absolute value amplitude of the long range potential. 

\textbf{Escape from classical localization}. Each short range scattering event transfers electron orbit by typical distance $R_c$ resulting in the effective microscopic diffusion constant $D_{mic}\sim L_pR_c w/U_p$ Eq.(\ref{DiffD}), where $U_p$ and $L_p\gg R_c$ are the typical amplitude and length of the long range potential. The average over electrons macroscopic diffusion constant $D_{mac}$ is estimated as the hopping length $L_p$ squared from one 'localization lake' to adjacent 'localization lake' over the time it takes for electron to diffuse to the percolating level $t\sim L_p^2/D_{mic}$. Thus, $D_{mac}\sim D_{mic}$. This conclusion is wrong. Let us prove that electrons with drifting orbit close to the percolating level have much larger diffusion constant and dominate the average macroscopic diffusion constant. In the vicinity of percolating saddle point long range potential is approximated as 
\begin{equation}\label{percPot}
U_p(\vec{r})=U_p(x^2-y^2)/2L_p^2.
\end{equation}
We use parameterization $x=r\cosh(\phi)$ and $y=r\sinh(\phi)$ with Jacobian: $dxdy=rdrd\phi$. For $\phi\gg 1$ we average the scattering rate Eq.(\ref{AGtaures}) over those trajectories that are closer than distance $R_c$ to the percolating line:
\begin{equation}\label{ScEvent}
\langle \frac{1}{\tau}\rangle \sim \frac{1}{L_p^2}\int^{L_p} \frac{1}{\tau} \ \theta(R_c-re^{-\phi})\  rdrd\phi,
\end{equation}
where the rate $1/\tau$ is given by Eq.(\ref{AGtaures}) with the local field $|\vec{\partial}U_p|\sim U_p r e^\phi /L_p^2$. After scattering with the probability Eq.(\ref{ScEvent}) electron will hop onto adjacent trajectory that diverges from the old trajectory after passing the saddle point by distance $L_p$. Therefore we estimate $D_{mac}\sim \mathcal{L} L_p^2 w/U_p\gg D_{mic}$, where $\mathcal{L}= \log(L_p/R_c) \log(\sqrt{\pi n}R_c)$. 

\section{Diffusion and Ryzhii conductivity}

Quantum states of an electron driven by uniform microwave $ac$-field are explicitly time depend. There exists a unitary transformation into the oscillating reference frame where wave functions of electron become time independent whereas the disorder potential becomes time dependent. The correlation function of short range disorder potential reads:
\begin{equation}\label{OscVel}
S_w(t,t';\vec{q})=S_w(\vec{q})\ e^{\displaystyle i\vec{q}\left(\vec{R}(t)- \vec{R}(t')\right)},
\end{equation}
where $\vec{R}(t)=\left(X(t),Y(t)\right)$ is the classical elliptical trajectory of charged particle in crossed constant magnetic field and microwave electric $ac$-field $E(t)=E\cos(\omega t)$ with frequency $\omega$ and polarized linearly along $x$ axis:
\begin{eqnarray}
X(t)=\frac{\mathcal{E}l_H}{\sqrt{2N}} \frac{\omega_H\cos(\omega t)}{\omega^2-\omega_H^2} \nonumber\\
Y(t)=\frac{\mathcal{E}l_H}{\sqrt{2N}} \frac{\omega^2_H\sin(\omega t)}{\omega(\omega^2-\omega_H^2)}, 
\end{eqnarray}
where the amplitude of $ac$-field is conveniently expressed in the energy terms: $\mathcal{E}=eER_c$. For long range disorder we neglect the time-dependent phase in the correlation function Eq.(\ref{OscVel}) due to the small wavevector $\vec{q}$. For the short range disorder this phase is essential and we transform the correlation function further: 
\begin{equation}\label{DrivenLine}
S_w(t,t';\vec{q})=S_w(\vec{q}) \sum_{m=-\infty}^{\infty}  J^2_m\left(q\mathcal{R}\right) e^{im\omega (t-t')}, 
\end{equation}
where $J_m(x)$ is the Bessel function and the size of the elliptical trajectory is:
\begin{equation}
\mathcal{R}=\frac{\mathcal{E}l_H}{\sqrt{2N}}\frac{\omega_H\sqrt{\omega^2+\omega_H^2}}{\omega (\omega^2-\omega_H^2)}
\end{equation} 
First we take average over the short range disorder for a given realization of the long-range potential. This calculation follows identically that of Ref.\cite{kash04}:
\begin{eqnarray} \label{Conductivity}
\sigma_{xx}(\Omega)\pm\sigma_{yy}(\Omega)=\frac{4e^2N}{\pi} w\sum_m\int \frac{d\epsilon}{2\pi}\left\{ \begin{array}{c} P_m \\ S_m \end{array} \right\} \nonumber\\ \frac{f(\epsilon)-f(\epsilon+m\omega-\Omega)}{\Omega} \textrm{Im} G(\epsilon) \textrm{Im} G(\epsilon+m\omega-\Omega)
\end{eqnarray}
where the axis $x$ is along and the axis $y$ is perpendicular to the $ac$-electric field and
\begin{eqnarray}\label{iml}
S_m\! =\!\! \int\! J^2_m(q\mathcal{R}) \left[J_0(qR_c)J_2(qR_c)+ J^2_1(qR_c) \right] \frac{d^2\vec{q}}{(2\pi)^2} \nonumber\\
P_m= \int J^2_m(q\mathcal{R}) \left[J^2_0(qR_c)- J^2_1(qR_c) \right] \frac{d^2\vec{q}}{(2\pi)^2}.
\end{eqnarray}
Next we average Eq.(\ref{Conductivity}) over the long-range disorder potential using the method of the previous section. In the limit $l_H\ll 2d\ll R_c$ each Green function in Eq.(\ref{Conductivity}) is averaged separately. The $dc$-conductivity $\Omega=0$ reads:
\begin{eqnarray} \label{ConductDC}
\sigma_{xx}\pm\sigma_{yy}=\frac{4e^2N}{\pi}w\sum_m\int \frac{d\epsilon}{2\pi} \left\{ \begin{array}{c} P_m \\ S_m \end{array} \right\} \textrm{Im} G(\epsilon+m\omega) \nonumber\\ \left[(f(\epsilon+m\omega)-f(\epsilon))\frac{d}{d\epsilon}\textrm{Im} G(\epsilon)- \frac{df}{d\epsilon}\textrm{Im} G(\epsilon) \right]
\end{eqnarray}
where the Green functions are averaged. Apart from the conductivity anisotropy and the transport corrections in our $P_m$ and $S_m$ Eq.(\ref{ConductDC}) coincides with the $dc$-conductivity found in Ref.\cite{ll03}. 

\begin{figure}
\includegraphics{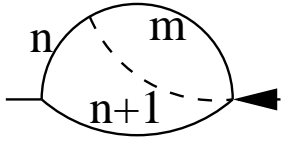}
\caption{Diagram for the average current in the left vertex. It is a linear response to the inhomogeneous electron density with the density gradient being indicated by the arrow.}
\end{figure}

\textbf{Diffusion constant in driven state.} We use the Keldysh method to calculate the current in the non-equilibrium state of electron system driven by microwaves with long-range gradient of electron density. The current is averaged during the time much shorter then the long range relaxation time. The local current density reads:
\begin{equation}\label{CurrentKeld}
\vec{j}(\vec{r},t)=\hat{\vec{j}}G^{+-}(\vec{r}t;\vec{r}t).
\end{equation}
Diffusion has an origin in the short range disorder scattering therefore we expand the Green function in Eq.(\ref{CurrentKeld}) in series over 'short range' impurity lines. One of the Green functions in each diagram of such expansion represents a variation of the occupation number distribution $\delta f(\epsilon)$:
\begin{equation}
\delta G^{+-}=\delta fG^A-G^R\delta f,
\end{equation}
whereas all Green functions to the left are retarded and all Green functions to the right are advanced. They are summed up into full Green functions and only one 'short range' impurity line remains. In weakly non-equilibrium Fermi liquid the variation of the occupation number distribution is localized in the energy domain in the vicinity of the Fermi level. Therefore we use ansatz:
\begin{equation}\label{deltaf}
\delta f(\epsilon)= -\frac{1}{\nu(\epsilon_F)}\frac{df}{d\epsilon}n_{\vec{q}} \exp(i\vec{q}\vec{r})
\end{equation}
It has a quasi-classical property: $\int \delta f(\vec{p})\ d^2\vec{p}/(2\pi)^2= n_{\vec{q}} \exp(i\vec{q}\vec{r})$. In the perpendicular magnetic field we expand $\exp(i\vec{q}\vec{r})= 1+iml_H^2\vec{q}\times\hat{\vec{j}}+O(q^2)$ in series of small $\vec{q}$, neglecting the magnetic shift operator that does not contribute to the current. A typical diagram that appears in the result is shown in Fig.3. It is proportional to $\vec{q}n_{\vec{q}}$ and therefore gives an additive to the linear response diffusion equation: $\vec{j}=D\vec{\nabla} n$. The dashed line in the microwave driven state is given by Eq.(\ref{DrivenLine}). The analytical expression for the sum of all such diagrams ($n=m$) gives the diffusion constant:
\begin{eqnarray}\label{diffusion}
D_{xx}\pm D_{yy}=-\frac{4N}{\pi\nu(\epsilon_F)}w\sum_m \int \frac{d\epsilon}{2\pi} \left\{ \begin{array}{c} P_m(\omega) \\ S_m(\omega) \end{array} \right\} \nonumber\\ \frac{df}{d\epsilon}\  \textrm{Im} G(\epsilon) \textrm{Im} G(\epsilon+m\omega).
\end{eqnarray}
Because $\textrm{Im}G(\epsilon)$ is positively defined function - it represents the density of states - the diffusion constant is positive $D_{aa}>0$. It coincides with the second term in the second line of the total conductivity Eq.(\ref{Conductivity}) that we call the diffusion conductivity $\sigma^D$. By definition they are related by the Einstein formula: $\sigma^D_{aa}=e^2\nu(\epsilon_F)D_{aa}$. But the Einstein relationship between the total conductivity and the diffusion constant is violated in the systems driven by microwaves due to the Ryzhii conductivity (the first term in the second line of the total conductivity Eq.(\ref{Conductivity})):
\begin{eqnarray}
\sigma^R_{xx}\pm \sigma^R_{yy}=\frac{4e^2N}{\pi}w\sum_m \int \frac{d\epsilon}{2\pi}\left\{ \begin{array}{c} P_m(\omega) \\ S_m(\omega) \end{array} \right\} \nonumber\\ \left(f(\epsilon)-f(\epsilon+m\omega)\right) \textrm{Im} G(\epsilon) \frac{d}{d\epsilon}\textrm{Im} G(\epsilon+m\omega)
\end{eqnarray}
Therefore we have proved that the total $dc$-conductivity Eq.(\ref{Conductivity}) is the sum of the diffusion conductivity and the Ryzhii conductivity: $\sigma(W)=\sigma_D(W)+ \sigma_R(W)$, for any intensity of the microwave $ac$-field.

\textbf{How zero resistance states emerge.} When the power of the microwave excitation provided $k\omega_H< \omega<(k+1/2) \omega_H$ is increasing the electron density distribution (originally due to the donor potential) become more and more even because the negative Ryzhii current moves electrons from the places of lower electrostatic energy (higher density) to places of higher electrostatic energy (lower density). The correlation function of screened donor potential is changing from the completely screened Eq.(\ref{CorVV}) to the unscreened Eq.(\ref{SpacerIm}). And the smaller is the wavevector the more prominent in Eq.(\ref{SpacerIm}) a harmonic of the donor potential becomes. The average harmonic magnitude is $q$ dependent Eq.(\ref{screening}):
\begin{equation}\label{harm}
u(\vec{q},W)=\frac{2\pi e^2}{|\vec{q}|+2\pi e^2\nu(\epsilon_F) (\sigma^D(W)-\sigma^R(W))/\sigma^D(W)}
\end{equation} 
We conclude that the harmonic with $\vec{q}=0$ will diverge first as the microwave intensity and $\sigma^R(W)$ both grow. The magnitude of the long range potential grows to infinity at the critical microwave power $W_c$: $\sigma^R(W_c)= \sigma^D(W_c)$. At powers higher than the critical $W>W_c$ the harmonics in Eq.(\ref{harm}) that belong to the shell $|\vec{q}_c|= \alpha(W-W_c)$ are divergent and therefore the local electric field will become infinite as well as the amplitude. The inelastic electron relaxation due to phonons will probably prevent the gradient of the long range potential to grow to infinity. $q_c$ could be considered as an order parameter of an emergent large gradient long range potential. Due to the obvious frustration in the electron system, that should handle large amplitude harmonics with ever changing wavevectors as the power grows, some non-equilibrium glass-like pattern of long range potential will probably freeze out eventually. In this situation the diffusion is dominated by electrons drifting in very long range ($qR_c\ll 1$) random potential with very large local electric fields. As discussed in the previous Section the diffusion constant will be very small - inversely proportional to the magnitude of the electric field on the percolating contour. Therefore the conductivity is almost zero and a zero-resistance state will emerge.

On the other hand the Hall conductivity in 2DEG with almost even electron density and in the long range potential is determined by the fast drift (without closures) by electrons that are close to the percolating level in the direction perpendicular to the applied Hall electric field. This mechanism gives the Hall conductivity $\sigma_{xy}=enc/H$, with $n$ being the electron density on the percolating energy level. 

\section{Conclusion}

For 2DEG in magnetic field under the microwave excitation we have found the effect of the Ryzhii current on the Coulomb screening of long range external potential. The screening factor is divergent at some critical microwave intensity. At this and higher powers very large amplitude harmonics of very long range potential develops. We have found the diffusion constant for quasiparticles in such large amplitude long range potential due to the scattering on weak short range disorder. The larger is the long range potential the smaller is the longitudinal conductivity - therefore the states with almost zero resistance emerge. 

\begin{acknowledgments}
I would like to thank the Grenoble high magnetic fields laboratory for hospitality and S.V. Iordanski for constant interest to this work and many useful discussions and suggestions. This work was supported by Russian fund for basic research under grants \# 03-02-17229-a and \# 05-02-16553-a.
\end{acknowledgments}


\begin{thebibliography}{22}
\bibitem{bla-bla} There is a gauge that a typical wave function of the $N$-th Landau level has a size $\sim R_c$ and has $\sim N$ zero value lines because it has to be orthogonal to all lower Landau level wave functions. The area per state is $2\pi l_H^2$, therefore inside a circle of size $R_c$ there are many states ($\sim N\gg 1$) of the $N$-th Landau level- we call it local degeneracy. The classical motion of quasiparticle in long range potential is a precession in circle of radius $R_c$ and a smooth drift of the center along the line of the constant energy level \cite{PG}. If a short range Gaussian disorder is present then the local degeneracy is lifted into a short range energy band and the classical trajectory becomes random. In strong long range fields quasiparticle still drifts along the energy level but after shifting by distance $\sim R_c$ it could change the state from the local degeneracy and therefore change the short range energy inside the local short range energy band. Because the total energy is conserved quasiparticle will skip onto an adjacent energy level of the long range potential. Thus a diffusion inside a band occurs.
\end{thebibliography}
\end{document}